\documentclass[]{pasj01}
\draft

\begin{document} 
\Received{}%{yyyy/mm/dd}
\Accepted{}%{yyyy/mm/dd}
%\Published{yyyy/mm/dd}

\title{Machine Learning Based Real Bogus System for HSC-SSP Moving Object Detecting Pipeline}

%%% begin:list of authors
% Do NOT capitalize all letters in "textsc".
\author{Hsing-Wen \textsc{Lin}\altaffilmark{1}}%
%\thanks{Example: Present Address is xxxxxxxxxx}}
\altaffiltext{1}{Institute of Astronomy, National Central University, 32001, Taiwan}
\email{edlin@gm.astro.ncu.edu.tw}

\author{Ying-Tung \textsc{Chen}\altaffilmark{2}}
\altaffiltext{2}{Institute of Astronomy and Astrophysics, Academia Sinica, P. O. Box 23-141, Taipei 106, Taiwan}
\email{ytchen@asiaa.sinica.edu.tw}

\author{Jen-Hung \textsc{Wang}\altaffilmark{2}}

\author{Shiang-Yu \textsc{Wang}\altaffilmark{2}}

\author{Fumi \textsc{Yoshida}\altaffilmark{3,4}}
\altaffiltext{3}{Planetary Exploration Research Center, Chiba Institute of Technology, 2-17-1 Tsudanuma,Narashino, Chiba 275-0016, Japan}
\altaffiltext{4}{Department of Planetology, Graduate School of Science, Kobe University, Kobe, 657-8501, Japan}

\author{Wing-Huen \textsc{Ip}\altaffilmark{1}}

\author{Satoshi Miyazaki\altaffilmark{5, 6}}
\altaffiltext{5}{National Astronomical Observatory of Japan, 2-21-1 Osawa, Mitaka, Tokyo 181-8588, Japan}
\altaffiltext{6}{SOKENDAI(The Graduate University for Advanced Studies), Mitaka, Tokyo, 181-8588, Japan}
%\author{Yutaka Komiyama\altaffilmark{3}}
%\author{Satoshi Kawanomoto\altaffilmark{3}}
%\author{Hidehiko Nakaya\altaffilmark{3}}
%\author{Yukiko Kamata\altaffilmark{3}}
%\author{Hiroshi Karoji\altaffilmark{3}}
%\author{Yoshiyuki Doi\altaffilmark{3}}
%\author{Hisanori Furusawa\altaffilmark{3}}
%\author{Tomoki Morokuma\altaffilmark{4,10}}\altaffiltext{4}{Institute of Astronomy, University of Tokyo, 2-21-1 Osawa, Mitaka, Tokyo 181-0015,Japan}
%\author{Kyoji Nariai\altaffilmark{3}}
%\author{Yoko Tanaka\altaffilmark{3}}
%\author{Fumihiro Uraguchi\altaffilmark{3}}
%\author{Yousuke Utsumi\altaffilmark{5}}
%\altaffiltext{5}{Hiroshima Astrophysical Science Center, Hiroshima University, 1-3-1 Kagamiyama, Higashi-Hiroshima, Hiroshima, 739-8526, Japan}
%\author{Yoshiyuki Obuchi\altaffilmark{3}}
%\author{Tadafumi Takata\altaffilmark{3}}
%\author{Yuki Okura\altaffilmark{6,7}}
%\altaffiltext{6}{RIKEN High Energy Astrophysics Laboratory,2-1 Hirosawa, Wako, Saitama 351-0198, Japan}
%\altaffiltext{7}{RIKEN BNL Research Center, Bldg. 510A, 20 Pennsylvania Street, Brookhaven National Laboratory, Upton, NY 11973}
%\author{Yoshihiko Yamada\altaffilmark{3}}
%\author{Hitomi Yamanoi\altaffilmark{3}}
%\author{Michitaro Koike\altaffilmark{3}}
%\author{Daigo Tomono\altaffilmark{3}}
%\author{Tomio Kurakami\altaffilmark{3}}
%\author{Kazuto Namikawa\altaffilmark{3}}
%\author{Tomonori Usuda\altaffilmark{3}}
%\author{Fumiaki Nakata\altaffilmark{3}}
\author{Tsuyoshi Terai\altaffilmark{7}}
\altaffiltext{7}{Subaru Telescope, National Astronomical Observatory of Japan, 650 North A`ohoku Place, Hilo, HI 96720, USA}
 %Do NOT move this preamble from here!

\maketitle

\begin{abstract}
Machine learning techniques are widely applied in many modern optical sky surveys,  e.q. Pan-STARRS1, PTF/iPTF and Subaru/Hyper Suprime-Cam survey, to reduce human intervention for data verification. In this study, we have established a machine learning based real-bogus system to reject the false detections in the Subaru/Hyper-Suprime-Cam Strategic Survey Program (HSC-SSP) source catalog. Therefore the HSC-SSP moving object detection pipeline can operate more effectively due to the reduction of false positives. To train the real-bogus system, we use the stationary sources as the real training set and the `flagged' data as the bogus set. The training set contains 47 features, most of which are photometric measurements and shape moments generated from the HSC image reduction pipeline ({\tt hscPipe}). Our system can reach a true positive rate (tpr) $\sim 96\%$ with a false positive rate (fpr) $\sim 1\%$ or tpr $\sim 99\%$ at fpr $\sim 5\%$. Therefore we conclude that the stationary sources are decent real training samples, and using photometry measurements and shape moments can reject the false positives effectively. 
\end{abstract}

\section{Introduction}
Synoptic optical sky surveys can generate huge amounts of imaging data every night and detect billions of astronomical sources. Along with these scientifically valuable sources, such as stars, galaxies, supernovae, asteroids and etc., there are significant non-astrophysical detections in the data. It requires humans to inspect the images and look for new events. 

Current modern sky survey data-sets are too large to be examined only by humans without any computer's intervention. Therefore, the machine learning (ML) techniques to automatically identify and classify astronomical sources have been introduced in many survey projects, e.q. Pan-STARRS1, PTF/iPTF and Dark Energy Survey. There are many applications for the ML in astronomy, for examples, 1. to identify the real astronomical sources from many of artifacts  (real-bogus system): The process of image differencing introduces many of the artifacts that need to be removed using the ML software. \citep{bai07, bri13, blo12, gol15, mas17, mor16, wri15}, 2. to classify the astronomical sources using multi-color photometry, spectrum \citep{bar17, blo12, bui15, mil17} and/or lightcurves \citep{hup17, mil15}, 3. to estimate the photometry red-shift \citep{cav17, cav15, ger10, kro14, sad16, sam17, wol17, zhe12}, and 4. to identify fast-moving near-Earth asteroids \citep{was17}.

The Subaru/Hyper-Suprime-Cam (HSC) \citep{kaw17, kom17, miy17} Strategic Survey Program (SSP) \citep{arm17, bos17, tak17} is one of the current largest optical surveys. It is planned to be executed over 6 years from 2014, to mainly investigate the mystery of dark matter/ dark energy as well as galaxy evolution, high red-shift AGNs and other kinds of astrophysical topics. Undoubtedly, this survey will suffer the influences of many false positives. To conquer the problem, \citet{mor16} built a real-bogus system by training three different kinds of supervised learning (AUC Boosting, Random Forest, and Deep Neural Network) with artificial objects. The system has been applied to select true optical transients in the differential images of the two HSC/SSP ultra-deep fields: COSMOS and SXDS, and it is able to reduce the false positive rate to $ 1.0\%$, and reaches the true positive rate of $90\%$ in the magnitude range from 22.0 to 25.0 magnitudes.  

This result looks promising. However, like all other real-bogus systems, this machine works for the differential images (the single epoch ``new'' images have a ``reference image'' subtracted. The reference images is the deep co-add of many exposures.) and is unsuitable in some cases. For example, if we need the real-time process of the survey data, the difference images may be unavailable or not available in time. Searching for the solar system moving objects in HSC-SSP wide survey is one of the cases; the reference stacked images of HSC-SSP wide fields will not be made until the survey data is released, which has the timescale about a year. However, even the slow moving Trans-Neptune objects (TNOs) need to be tracked within a few months. Therefore the moving object detecting pipeline has been developed for searching the moving objects with single exposure, non-differential source catalogs. The detailed algorithm of the pipeline is described in \citet{che17}.

Basically, the moving object pipeline detects the moving objects by searching for sequences of detections within a few nights. Therefore, the total time span to finish the search is proportional to the number of sequence detections. It is roughly proportional to the square of the total non-stationary sources.  
Although the false positives should be randomly distributed, they still have the chances to align in a row and become a moving object-like sequences of detections. As a result, the false positives can greatly decrease the searching speed by producing too many possible false sequences, and how to reduce the false positive in the source catalogs is a major task to complete the moving object search within reasonable computational time.  

In this study, we seek for an alternative real-bogus system that can work for the normal, non-differential source catalogs. In addition, to make this system easy to access, we will only use the features provided in HSC image reduction pipeline ({\tt hscPipe}) and avoid measuring extra moments of the objects from the images. This paper is structured as follows. In section 2, we explain how to collect the training data set and select the features. In section 3, we introduce the ML methods of the real-bogus system. In section 4 we evaluate the efficiency of the ML real-bogus system. The discussion and summary are in section 5 and section 6, respectively.

%\noindent IMPORTANT NOTICE\\
%1. ``\verb|\draft|'' creates single column and double spaces format.\\
%2. If you comment out ``\verb|\draft|'', the output will be double column
%   and single space.\\
%3. For cross-references, the use of ``\verb|\label|, \verb|\ref|, \verb|\cite|'' 
%   and the thebibliography environment is strongly recommended. \\
%4. Do NOT use ``\verb|\def|, \verb|\renewcommand|''.\\
%5. Do NOT redefine commands provided by PASJ01.cls.\\

\section{Training Set and Feature Selections}
A key point for building an effective, unbiased ML classifier is to collect an accurate, unbiased, and large training set. However, it is difficult in most cases. For example, human examination can provide accurate training samples, but it is time consuming, and as a result can lead to limited sample sizes. By contrast, selecting the training sample with some characteristics is very effective, but it could produce the biased and inaccurate samples. To conquer this problem, \citet{mor16} trained the ML classifier with synthetic training samples, which are both accurate and with large sample size. Still, the synthetic training set could be very biased, depending on the complexity of synthetic models. For example, it is relatively easier to make synthetic stars than synthetic galaxies, because galaxies have significantly more complex shapes and features.

Here we propose a simple way to collect the large training samples with high accuracy and small bias. The HSC data reduction pipeline ({\tt hscPipe}), provides lots of features in the source catalogs, including all kinds of photometry measurements, position measurements, and shape moments. The software team of {\tt hscPipe} has suggested some ``flags'' to filter the good and bad sources. However, it is possible that the ``good flags'' do not contain all of the faint real objects. On the other hand, the ``bad flags'' can be used for generating a set of ``pure garbages'', which is an ideal bogus training set for ML. The ``bad flags'' is consist with 1. {\tt centroid$\_$sdss$\_$flags == True} and {\tt flux$\_$aperture$\_$flags == True}: Both object centroiding and aperture flux measurement failed, or 2. {\tt flags$\_$pixel$\_$saturated$\_$any == True}: Any of the pixels in an object's footprint is saturated. For the detailed information of the HSC flag, please refer to \citet{arm17}.

To collect the pure real training set, we use the stationary sources to represent the real astronomical sources. Each stationary source is identified from the detections inside of a certain radius with the enough duration (time span) in a single SSP observation run. The stationary selection criteria are: \\
1. Search radius $\leq$ 0.5 arcsec: A stationary source has to have corresponding detections in different exposures within 0.5 arcsec radius. \\
2. Lowest number of detections $\geq$ 2: A stationary source must have at least two corresponding detections within the search radius. \\
3. Time span $\geq$ 20 minutes: The two corresponding detections have to be separated at least 20 minutes to avoid contamination from the slow-moving solar system objects, because the slow movers, such as distant TNOs, could have movement less than 0.5 arcsec within 20 minutes. \\
Here, we set the conditions of search radius and time span by calculating the excepted movements of a slow moving object at 100 AU at opposition.

Comparing the two different selection criteria of bogus and real detections, the bogus training set selection may not be very uniform, because the ``bad flags'' could cause some bias to select the artifacts, e.q. some optical flares could be successful centroided also without any saturated pixel; they can pass thought the ``bad flags'' filter.
On the other hand, the stationary sources collect almost all kinds of real detections with a whole range of brightness and shapes. These are more close to unbiased, uniform samples. Exception is the fast-moving asteroids, which have the streak-like shapes. Fortunately, in our case to search the distant moving objects, their moving rate is very slow and the shapes are very close to the normal stars. In short, we have better confidence in our real training set than the bogus training set.    

With the real and bogus training samples, the next step is the feature selection. Although the artifacts are very likely to be locational dependent, for example, they have a higher frequency to appear near the bright stars and the edges of CCD chips. Using the position dependent feature to select the artifacts will be very risky, since the ML may consider every source around the edges of CCD chips should be bogus. Thus, we avoid the use of location dependent features. The shape moments and photometry measurements are the ideal features to separate the real and bogus sources, not only because the artifacts and real sources could have different shapes or brightness, but also they are the location dependent features. Thus we mainly use the shape moments and photometry measurements to build our real-bogus system.  

Table~\ref{tab1} shows all of the features we selected from the HSC source catalogs. The Features are all related to the shape and photometry, except the ``{\tt parent}'' and ``{\tt flux$\_$aperture$\_$nInterpolatedPixel}''. There are four kinds of shape moments were selected in this study: {\tt shape$\_$sdss}, {\tt shape$\_$hsm$\_$moments}, {\tt shape$\_$sdss$\_$psf} and {\tt shape$\_$hsm$\_$psf$\_$moments}. {\tt shape$\_$sdss} is a reimplementation of the algorithm used in SDSS Photo Pipeline, while {\tt shape$\_$hsm$\_$moments} is a wrapper of the HSM \citep{hir03} and implementation in GalSim \citep{row15}. The {\tt shape$\_$sdss$\_$psf} and {\tt shape$\_$hsm$\_$psf$\_$moments} are the same algorithm with {\tt shape$\_$sdss} and  {\tt shape$\_$hsm$\_$moments}, respeatively, but measure the PSF model at the source position.

Both {\tt shape$\_$sdss} and {\tt shape$\_$hsm$\_$moments} are measured the shape with the adaptive moments. Adaptive moments are the second moments of the object intensity. They are measured by a particular scheme designed to have near-optimal signal-to-noise ratio. The three features provided by both algorithms are the sum of the second moments, the ellipticity (polarization) components and a fourth-order moment. For the detail of adaptive moments, we refer to \citet{ber02} and \citet{hir03}.

For the photometry measurements, there are ten different size of apertures (radius = 3, 4.5, 6, 9, 12, 17, 25, 35, 50, 70 pixels, respectively) implied in the {\tt hscPipe} for the aperture photometry. Considering the typical seeing size is about 3 pixel, we only selected the apertures with radius = 3, 4.5, 6, 9, 12 and 17 pixels as the six features of the aperture flux measurements. The {\tt flux$\_$sinc} is the elliptical aperture photometry using sinc interpolation. Finally, the only two features of non-photometry and non-shape moments are {\tt parent} and {\tt flux$\_$aperture$\_$nInterpolatedPixel}. The {\tt parent} records the number of peaks within the footprint of the detections, and the {\tt flux$\_$aperture$\_$nInterpolatedPixel} is the number of interpolated pixels in apertures. For the detailed information of the {\tt hscPipe} measurements, please refer to \citet{bos17}.

\begin{table}
\tbl{List of features 
\label{tab1}}{%
\begin{tabular}{@{}ll@{\qquad}lc@{}}  
\hline\noalign{\vskip3pt} 
feature & description & \\ 
\hline\noalign{\vskip3pt} 
parent & the number of peaks within the footprint of the detection& \\
shape$\_$hsm$\_$moments (three features) &source adaptive moments from HSM & \\
shape$\_$hsm$\_$psf$\_$moments (three features) & PSF adaptive moments from HSM & \\
shape$\_$sdss (three features) & shape measured with SDSS adaptive moment algorithm&\\
shape$\_$sdss$\_$psf (three features) & adaptive moments of the PSF model at the position&\\
flux$\_$aperture (six features) & sum of pixels in six different sized apertures&\\
flux$\_$aperture$\_$err (six features) & uncertainty for flux$\_$apertures&\\
flux$\_$aperture$\_$nInterpolatedPixel (six features) & Number of interpolated pixels in apertures &\\
flux$\_$gaussian & linear fit to an elliptical Gaussian with shape parameters set by adaptive moments&\\
flux$\_$gaussian$\_$err & uncertainty for flux$\_$gaussian&\\
flux$\_$kron & Kron photometry&\\ %\\ %: photometry with aperture set to some multiple of radius determined within some multiple of the source size&\\
flux$\_$kron$\_$err & uncertainty for flux$\_$kron&\\
flux$\_$kron$\_$radius & Kron radius (sqrt(a*b))&\\
flux$\_$kron$\_$psfRadius & Radius of PSF&\\
flux$\_$psf & flux measured by a fit to the PSF model&\\
flux$\_$psf$\_$err & uncertainty for flux$\_$psf&\\
flux$\_$sinc & elliptical aperture photometry using sinc interpolation &\\
flux$\_$sinc$\_$err & uncertainty for flux$\_$sinc &\\
flux$\_$gaussian$\_$apcorr & aperture correction applied to flux$\_$gaussian&\\
flux$\_$gaussian$\_$apcorr$\_$err & error on aperture correction applied to flux$\_$gaussian$\_$appcorr&\\
flux$\_$kron$\_$apcorr & aperture correction applied to flux$\_$kron&\\
flux$\_$kron$\_$apcorr$\_$err & error on aperture correction applied to flux$\_$kron&\\
flux$\_$psf$\_$apcorr & aperture correction applied to flux$\_$psf&\\
flux$\_$psf$\_$apcorr$\_$err & error on aperture correction applied to flux$\_$psf&\\
\hline\noalign{\vskip3pt} 
\end{tabular}}
\end{table}

\section{Machine Learning Algorithms for Real-Bogus Classifier}

Basically, the Real-Bogus separation is a two-class classification problem, and many of the ``supervised learning'' algorithms could be potentially useful to deal with it. The supervised learning is teaching the computer by the known-result training data to learn a general rule of mapping the input data to the output results. For example, in our case, we provide a set of detections to train the computer. Every detections have their own features and been labeled as `real' or `bogus'. After the training, the computer learns a general rule of that how to distinguish the `real' and `bogus' by examining their features.

Alternatively, we may also treat the Real-Bogus separation as a clustering problem, because the real and false detections can be potentially clustered into two different groups. In the clustering problem, the groups are unknown before training and make this an unsupervised task. Unlike the supervised learning, the unsupervised learning do not need any known results during the training, and leaving the computer on its own to figure out the structure of the training data.  

Since we have an imbalanced training samples, which the real set is much larger and may be less biased than the bogus set, except the supervised learning, we also give the unsupervised learning a try to build our Real-Bogus Classifier. Here we selected Random Forest and Isolation Forest for the supervised and unsupervised learning, respectively. 

\subsection{Supervised Learning: Random Forest} \label{RF}
Random Forest, RF \citep{bre01}, has been proved to be a very effective classifier for many kinds of data. It is a supervised ensemble-method composed by multiple decision trees, which are trained by their own subsets of training data. The subsets of training data are selected by bootstrap sampling (random sampling with replacement). The decision will be made by the majority-voting of every trees to reduce the variance of the model predictions. 

RF can process the extremely high dimensional data and evaluate the importance of each feature, which is suitable for our data. It also tends to produce less over-fitting than single decision tree, although the over-fitting is still possible. 

Below is the short description of the RF algorithm: \\
1. Make a training sub-set by bootstrap sampling from the training set, which has total N detections, for growing the tree. Keep the unselected samples, a.k.a out-of-bag samples, for the model evaluation. \\
2. If there are M features, randomly select m features from M (m $<$ M) at each node. Evaluate the optimal way on these m to split the samples. The value of m is held constant while we grow the forest.\\
3. Each tree is grown to the largest extent possible without pruning. \\
4. Repeat Step 1, 2 and 3 to build a forest of decision trees. \\
For more detailed descriptions of RF, we refer the reader to \citet{bre01}.

\subsection{Unsupervised Learning: Isolation Forest}
Isolation Forest, IF, \citep{liu08}, is an unsupervised learning method to search for anomaly detections or outliers in huge data-sets. On one hand, IF is similar to the RF, which is also constructed by multiple trees. On the other hand, unlike the RF which uses bootstrap sampling to select the training subsets for the decision trees, IF uses simple random sampling (without replacement) sub-sets to train the individual trees. The trees in the IF are not the decision trees; they do not make any decisions and are called {\it isolation trees}. Our motivation to use IF is that we have better and larger real training set. Therefore if we provide only the real set, we expect that the IF classifier will figure out how real detections looks like and reject the bogus. Below is the short description of the IF algorithm: \\
1. Randomly pickup a feature as a node of the tree and randomly select a value (between maximum and minimum) of the feature to divide the training samples into two groups. \\
2. Repeat step 1 for each groups, until reach the stop criterion. The criterion can be either a. every samples have been isolated, or b. reach the bottom of the tree. The counts of splittings required to isolate a sample is equivalent to the path length from the root node to the terminating node. \\
3. Repeat step 2 to generate a forest of isolation trees. \\
4. The mean path length, which is averaged over the path length of every tree in the forest, is represented the normality of the training samples. If a data point produces significant longer path length than the mean path length, it is very likely an anomaly detection.\\
For the detailed algorithm of IF method, we refer the reader to \citet{liu08}.  

\section{Implementation and Evaluation of the Real-Bogus Classifiers}

We applied the RF and IF implementation in {\tt scikit-learn 0.18}\footnote{An open-source machine learning library for Python (http://scikit-learn.org)} \citep{ped11}. The number of trees ({\tt n$\_$estimators}) in the classifiers can be specified in both RF and IF. With several tests we found that the out-of-bag error reaches a constant when the {\tt n$\_$estimators} is larger than $\sim 140$ to $160$. Therefore we set {\tt n$\_$estimators} = 160 during the whole studies. Notice that the out-of-bag error is the mean prediction error evaluated by using out-of-bag samples (see section~\ref{RF}). 

We trained RF and IF machines by inputting the HSC-SSP data in June 2016. The data was processed using {\tt hscPipe version 4.0.5}.
First, we separated all data of the observing run in June 2016 into {\tt HEALPixs} \citep{gor05} using python package {\tt healpy 1.9.1}\footnote{An open-source Healpix tools package for Python (https://pypi.python.org/pypi/healpy)} with parameter {\tt nside = 32}
(npix = 12288, mean spacing = 1.83 (r = 0.9), area = 3.34) and {\tt nest = True}. Detections taken in all kind of filters, different weather and seeing conditions are all included. Then we pick the detections in one specific {\tt HEALPix}, 6499, which has central coordinate RA and DEC of 208.1 and -001.2, respectively, as the training samples. The training samples contain 7,265,881 and 182,218 real and bogus detections, respectively. Afterward, we randomly split the training samples into the training set and the test set with a ratio of 6:4. The test set is withheld from training the RF and IF models in order to assess their accuracy as an independent sample of sources. The RF classifier was trained by both the real and the bogus training sets. The IF classifier, which based on unsupervised IF algorithms, was only trained by the real training set. Notice that some detections have feature values of ``nan'', which are not readable by both of the classifiers, therefore we use value ``-999'' instead of ``nan''.

Both RF and IF classifiers were evaluated by the same test set, which includes both the real and bogus detections. To test the efficiency consistency of the systems in different sky location or different SSP observing runs, we built two extra test sets. One is using the data in {\tt HEALPix} 6505, centered on RA=206.7 and Dec=+000.0, to represent the data taken in the same observing run but different sky location. The other one is using the data in {\tt HEALPix} 6491 (RA=216.6 and Dec=+001.2), which was observed in May 2015, to represent the data taken in different SSP observing run.

\subsection{ROC curves}
The receiver operating characteristic curves (ROC curves) is created by plotting the true positive rate against the false positive rate at various threshold settings. The true positive rate (tpr) is the proportion of positives that are correctly identified as such. On the other hand, the false positive rate (fpr) is the proportion of negatives that are incorrectly identified as positives. The ROC curve is useful to demonstrate  performance of a binary classifier and select the optimal threshold for the data separation.

\begin{figure}
\includegraphics[width = 8cm]{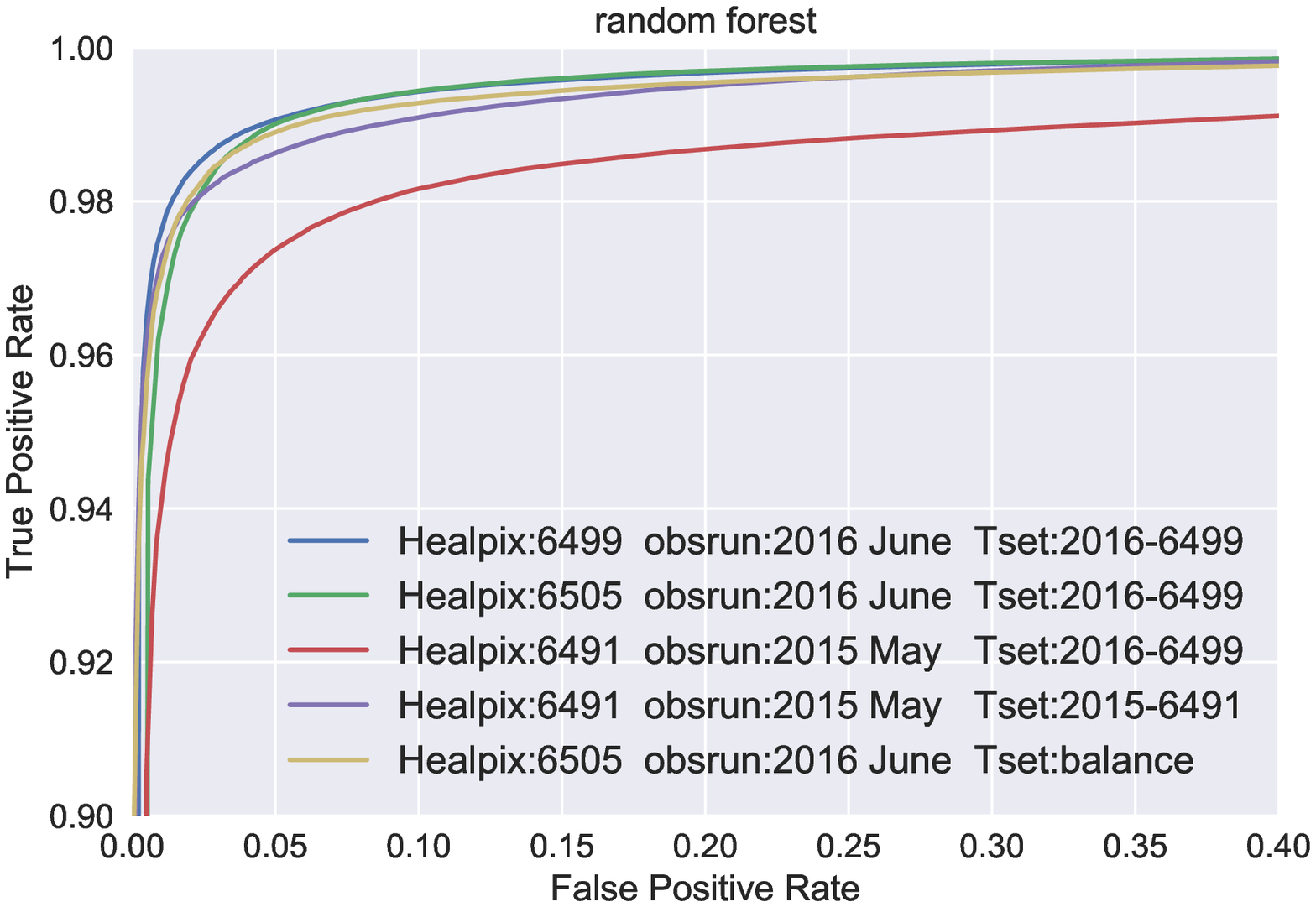} 
\includegraphics[width = 8cm]{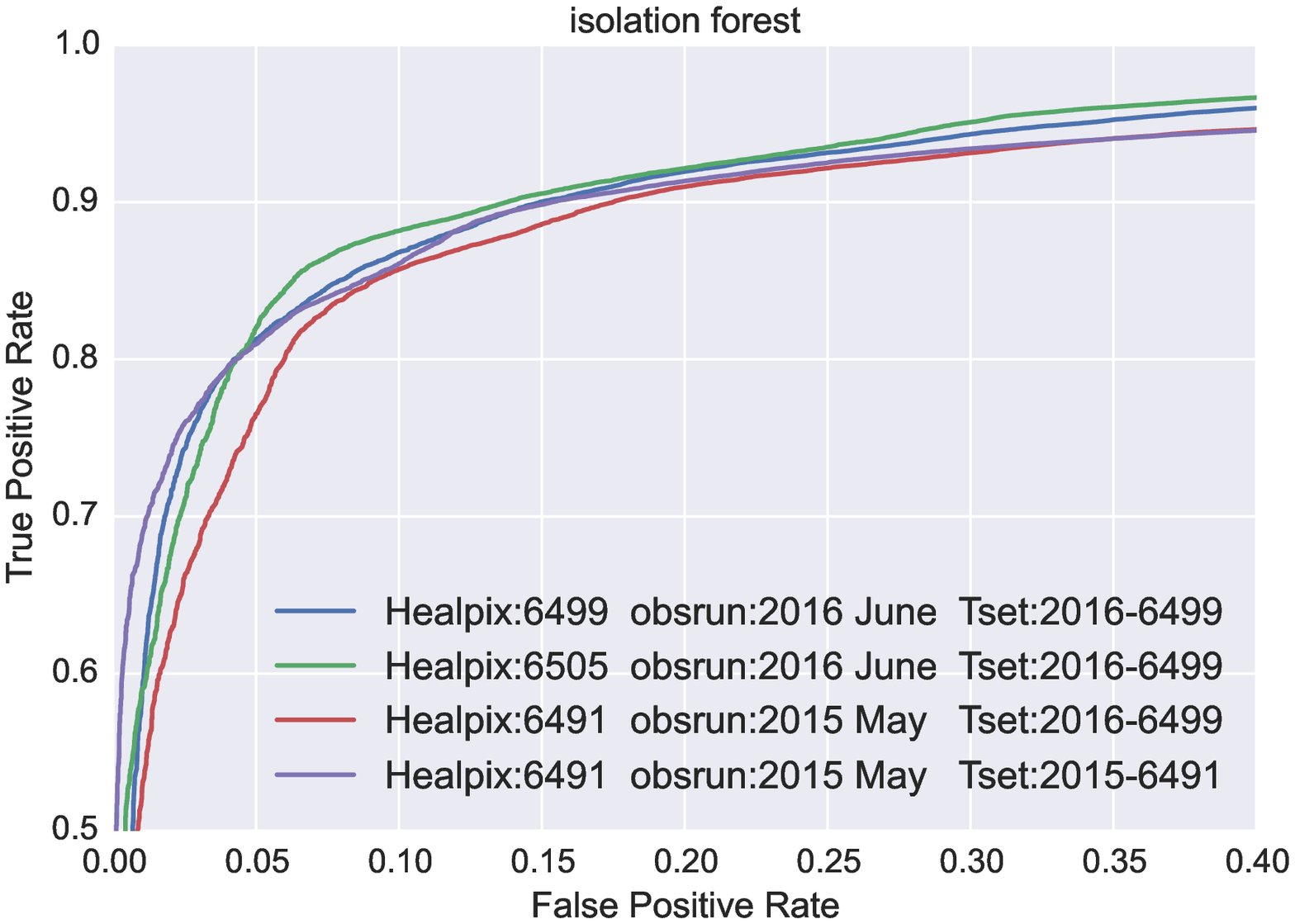} 
\caption{The ROC curves of left: RF, and right: IF classifier. The blue, green and red curves present the results of the same ML model apply in the same {\tt HEALPix} in the same observation run, different {\tt HEALPix} but the same observation run and different {\tt HEALPix} in different observation runs, respectively. The purple curve shows the result of re-training the ML model with the data of the case of red curve. The yellow curve in RF plot present the result of model that trained by balance training set.}\label{fig:1}
\end{figure}

The performance of our Real-Bogus system are shown in Figure~\ref{fig:1} as the ROC curves. For the RF classifier, it can achieve the tpr of $\sim 99\%$ at the point if the fpr $= 5\%$. If we only allow less than $1\%$ of fpr, this classifier can still provide more than $96\%$ of tpr. The performance is similar if we apply the same RF classifier to identify the detections in different {\tt HEALPix} but in the same observing run ({\tt HEALPix} 6505, the green curve). However, the performance decreases to tpr $\sim 97\%$ at fpr $= 5\%$ (red curve) if we apply the same classifier to evaluate the data of different SSP observing run ({\tt HEALPix} 6491). Nevertheless, this result is still very good. Finally, we re-train the RF classifier using the 2015 SSP observing run data, and the performance is back to normal ($\sim 99\%$ tpr at $5\%$ fpr, the purple curve). 

We noticed that the performance of the model on test data may be overstated due to the large imbalance training set. To test this issue, we random selected 2.5$\%$ of the real sample and made an balance training set with equal number of the ``real'' and ``bogus'' sources and retrained the model. The new model with balance training set perform almost the same ROC curve with the origin model with imbalance training set (yellow curve). We concluded that using the imbalance training set is appropriate in our case. 

On the other hand, the unsupervised IF classifier did not perform as well as RF classifier; it can only reach a tpr $\sim 90\%$ at a fpr $= 15\%$. The performance of IF classifier seems to be more independent to the different data; we tested it with the data-sets from both different {\tt HEALPix} and different SSP runs, and resulted in very similar the ROC curves. 

\subsection{Feature Distributions} \label{fd}

\begin{figure}
\includegraphics[width = .5\textwidth]{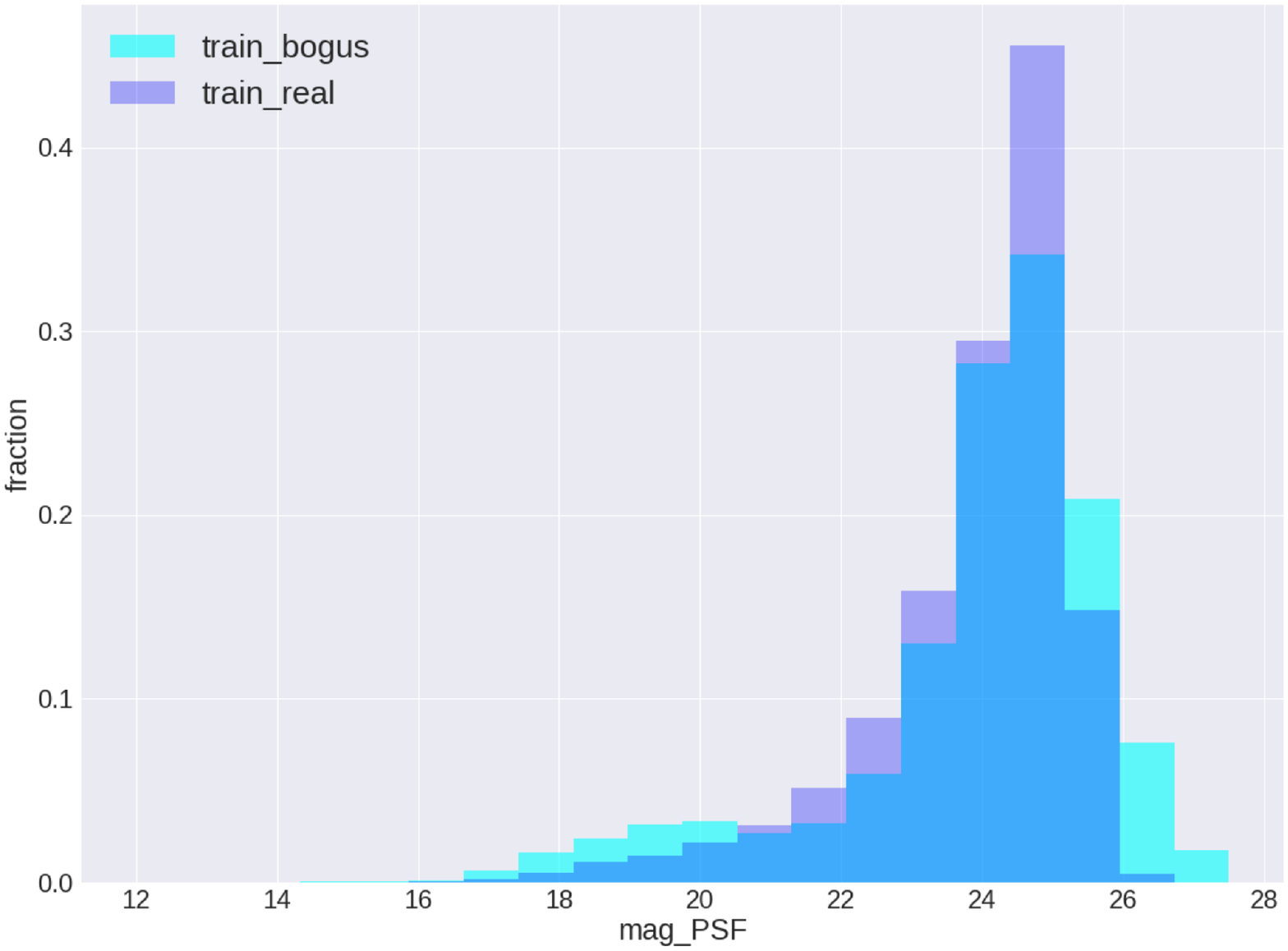} 
\includegraphics[width = .5\textwidth]{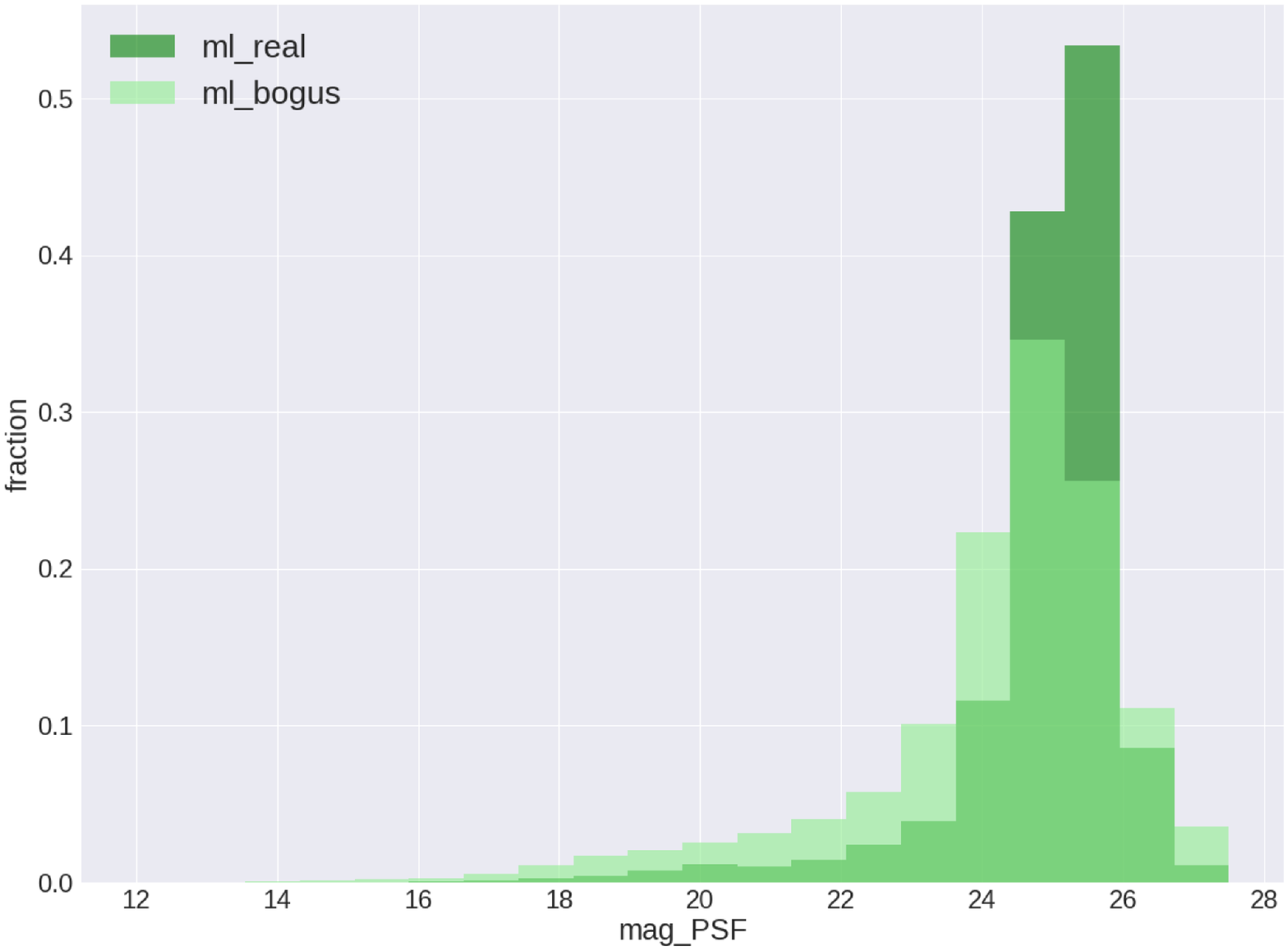} 
\caption{The PSF magnitude distribution of ``real'' and ``bogus'' sources for the training (left) and the ML selection samples (right).}\label{fig:5}
\end{figure}

Based on the previous result, we chose the RF classifier to build our real-bogus system. To understand the brightness dependency of the real-bogus separation, we first examine the magnitude distribution of ``real'' and ``bogus'' sources for both training and ML selection samples. The result show in Figure~\ref{fig:5}. The ``train$\_$real'' is represented the sample of stationary sources, and the ``train$\_$bogus'' is represented the ``bad flags'' sources. On the other hand, the ``ml$\_$real'' and the ``ml$\_$bogus'' are the non-stationary sources that were identified as ``real'' and ``bogus'' by the real-bogus system, respectively. Although the ``train$\_$real'' have fewer sources fainter than 26 magnitude, there is no evident that every non-stationary sources with brightness fainter than 26 are all belong to the ``ml$\_$bogus''. Furthermore, there is also no clear trend that ``ml$\_$bogus'' are generally fainter than ``ml$\_$real''. This result suggests that the real bogus separation is not the function of brightness, i.e. our system does not serve as a method of simple brightness cut.  

\begin{figure}
\includegraphics[width = .5\textwidth]{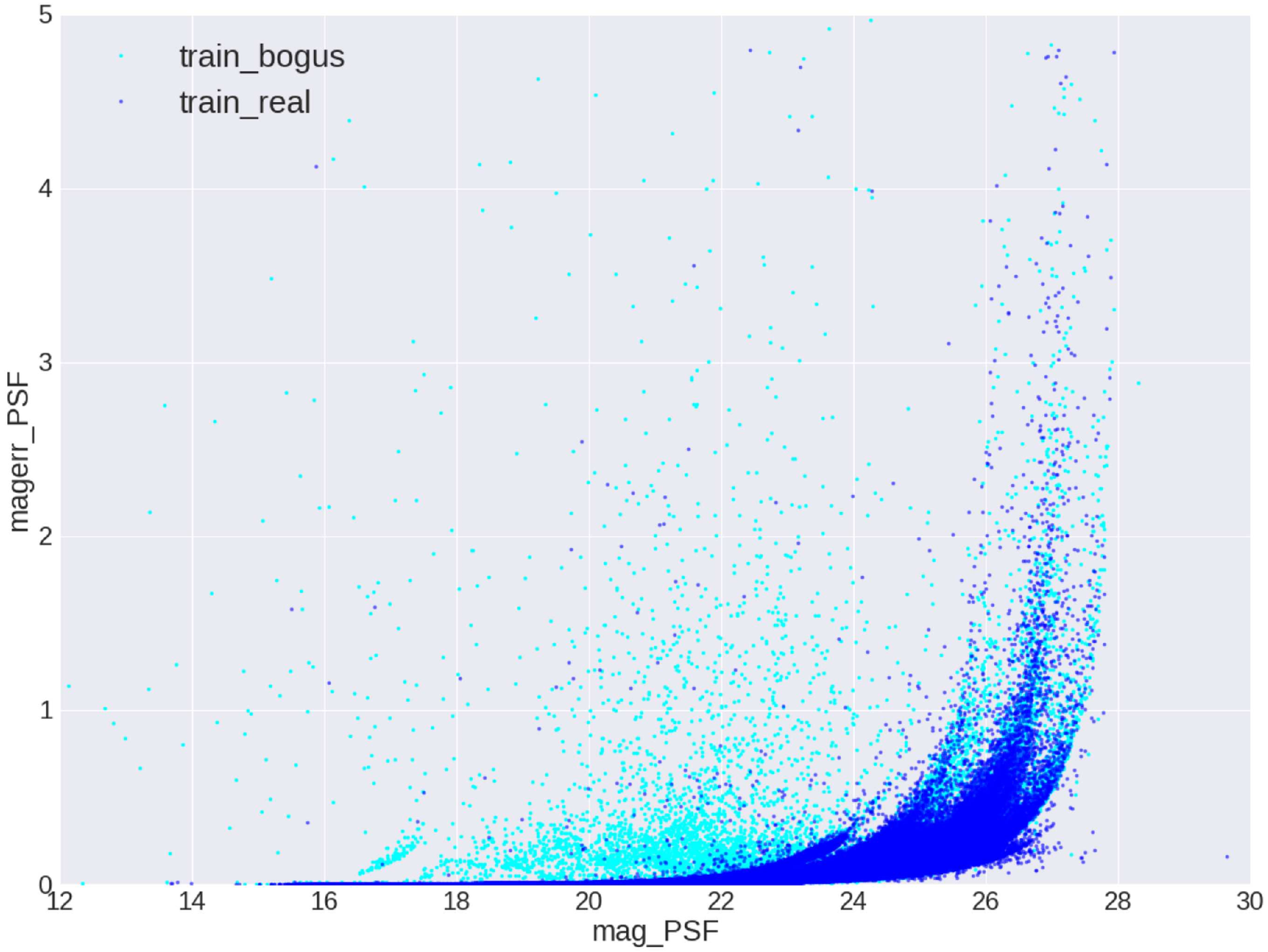} 
\includegraphics[width = .5\textwidth]{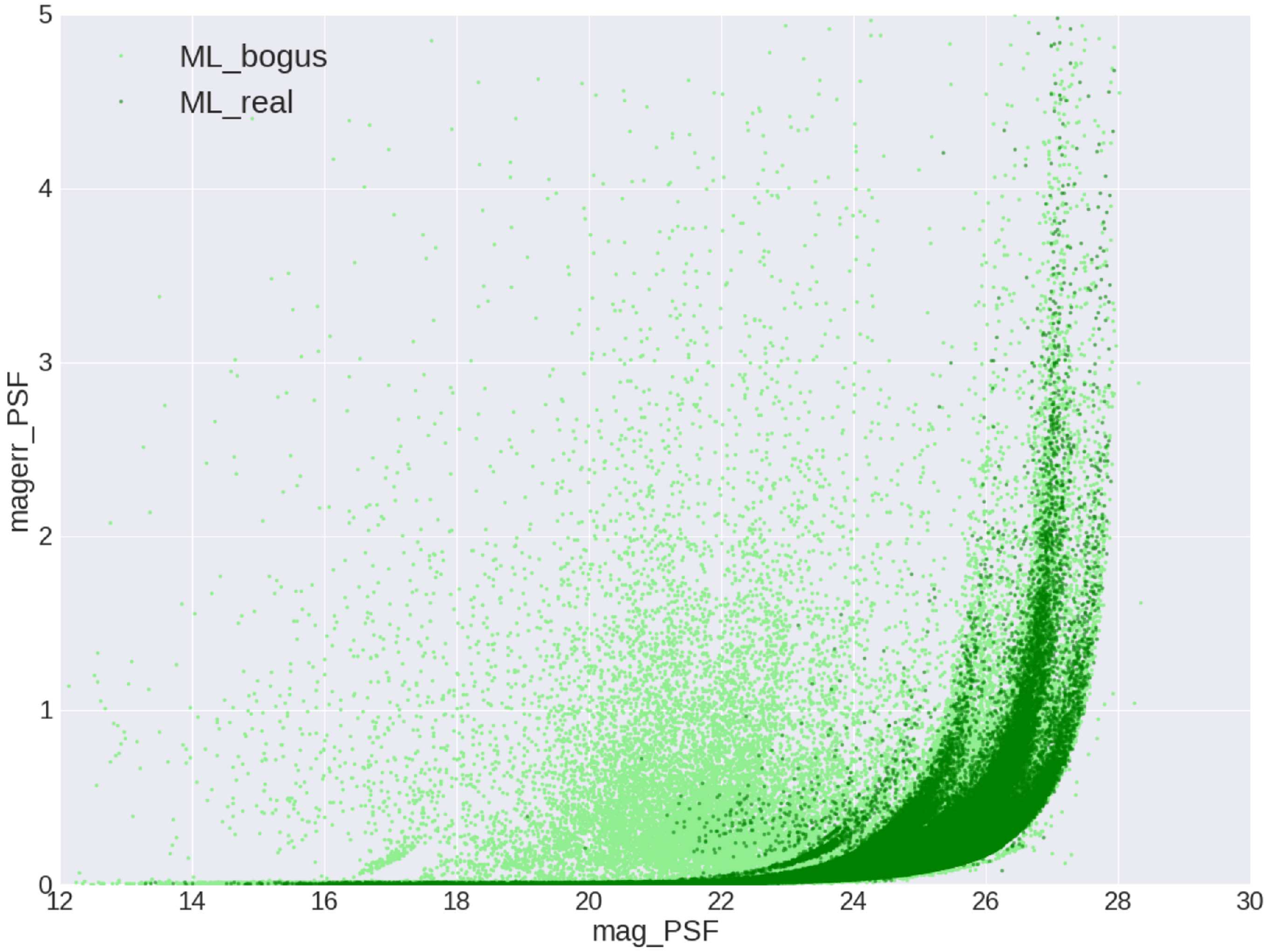} 
\caption{The PSF magnitude versus magnitude error of ``real'' and ``bogus'' sources for the training (left) and the ML selection samples (right).}\label{fig:6}
\end{figure}

Figure~\ref{fig:6} shows the distributions of magnitude versus magnitude error. The plots clear show that the real sources follow the reasonable brightness-uncertainly relation (the fainter sources have larger uncertainties) for both training and ML selection samples. By contrast, the bogus sources does not necessary follow this relation, and almost all of the out of track sources (the sources with unreasonable large uncertainty) have been identified as the ``bogus'' sources. Again, the plots show that our real-bogus system can select the reasonable sources and reject the abnormal ones, not just a simple brightness cut. We noticed that the major three different brightness-uncertainly relation curves in the plots should indicate the sources that were taken in three different filters.

\begin{figure}
\includegraphics[width = .5\textwidth]{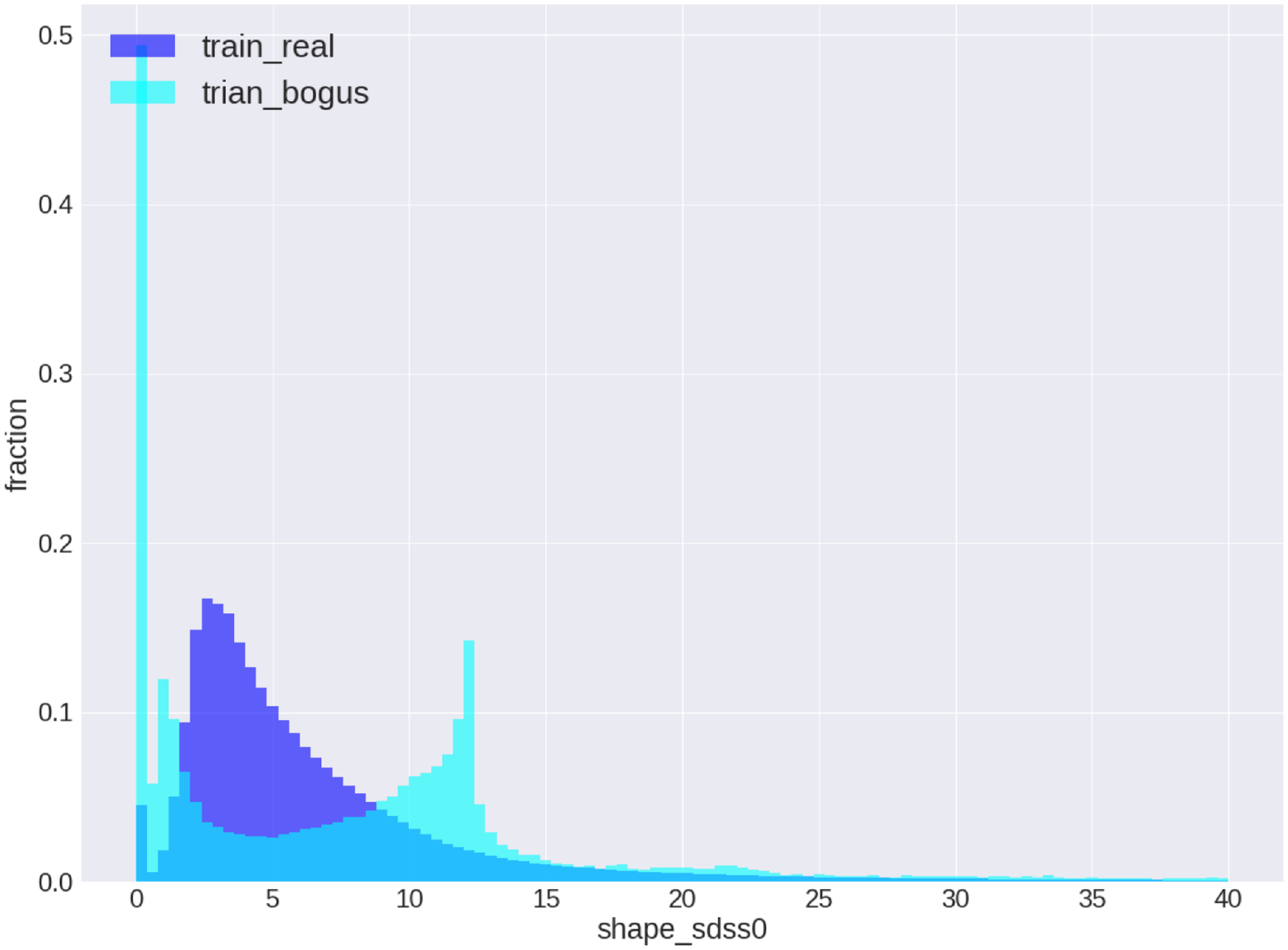} 
\includegraphics[width = .5\textwidth]{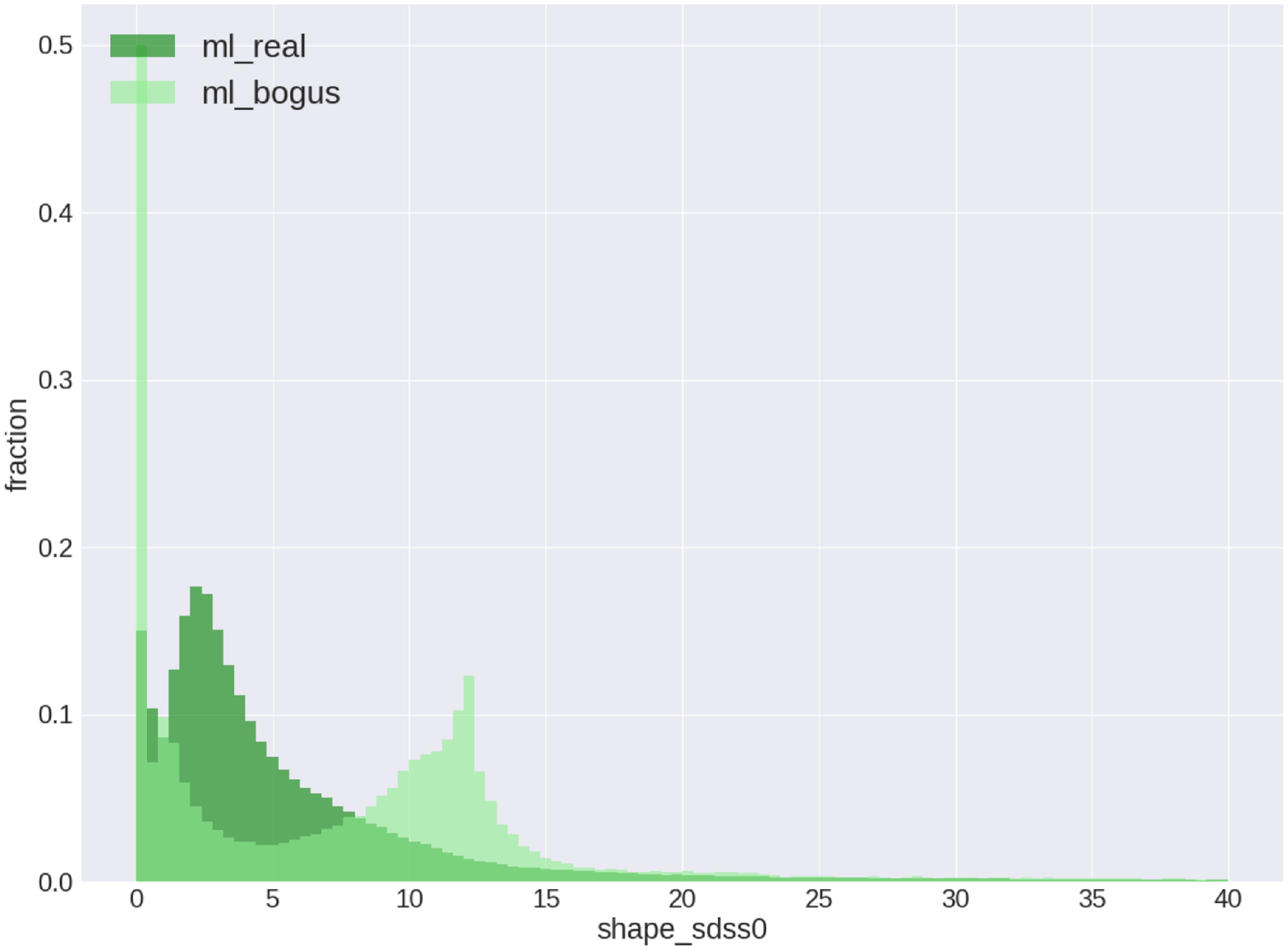} 
\includegraphics[width = .5\textwidth]{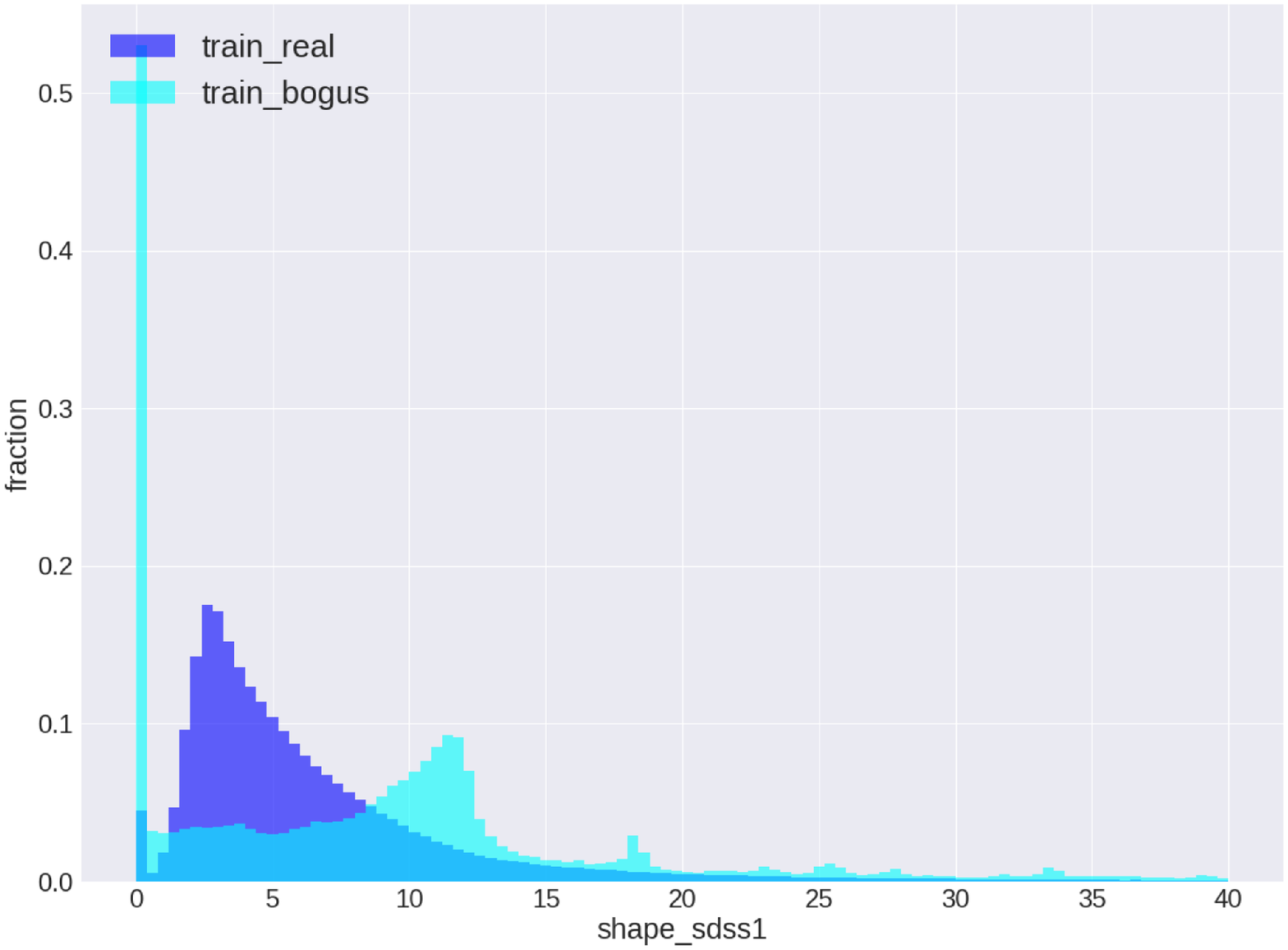} 
\includegraphics[width = .5\textwidth]{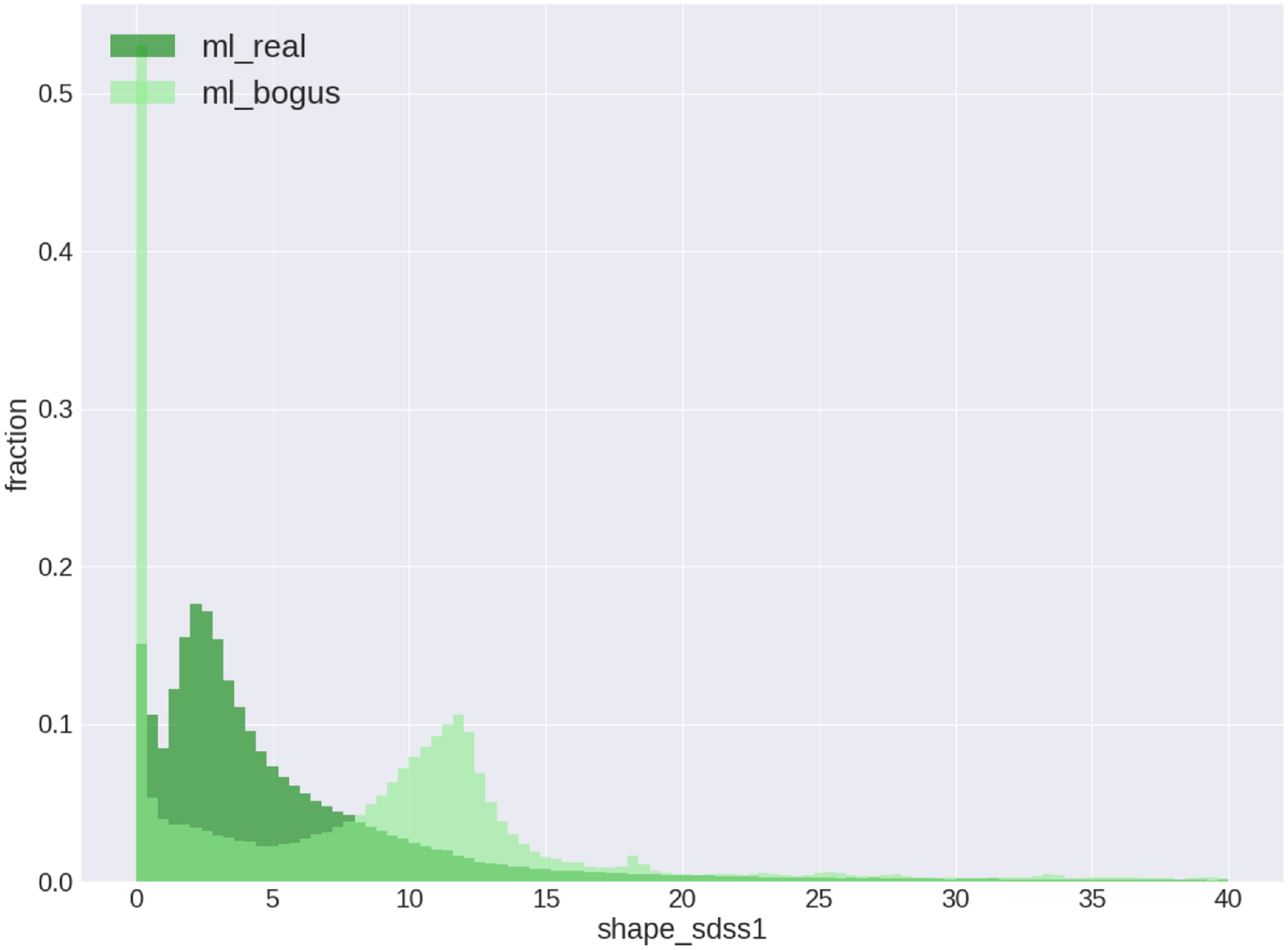} 
\caption{The shape moment distributions of ``real'' and ``bogus'' sources for the training (blue) and the ML selection samples (green). The shape$\_$sdss0 is the sum of the second moments, and shape$\_$sdss1 is the ellipticity components.}\label{fig:shape}
\end{figure}

Figure~\ref{fig:shape} shows the distributions of the shape moments. The plots clear show that ``real'' and ``bogus'' sources have different distributions in the training samples (blue), and our real-bogus selection results represent such distributions for both ML selected real and bogus detections (green).

\subsection{Visual Examination} \label{ve}

\begin{figure}
\includegraphics[width = 1.\textwidth]{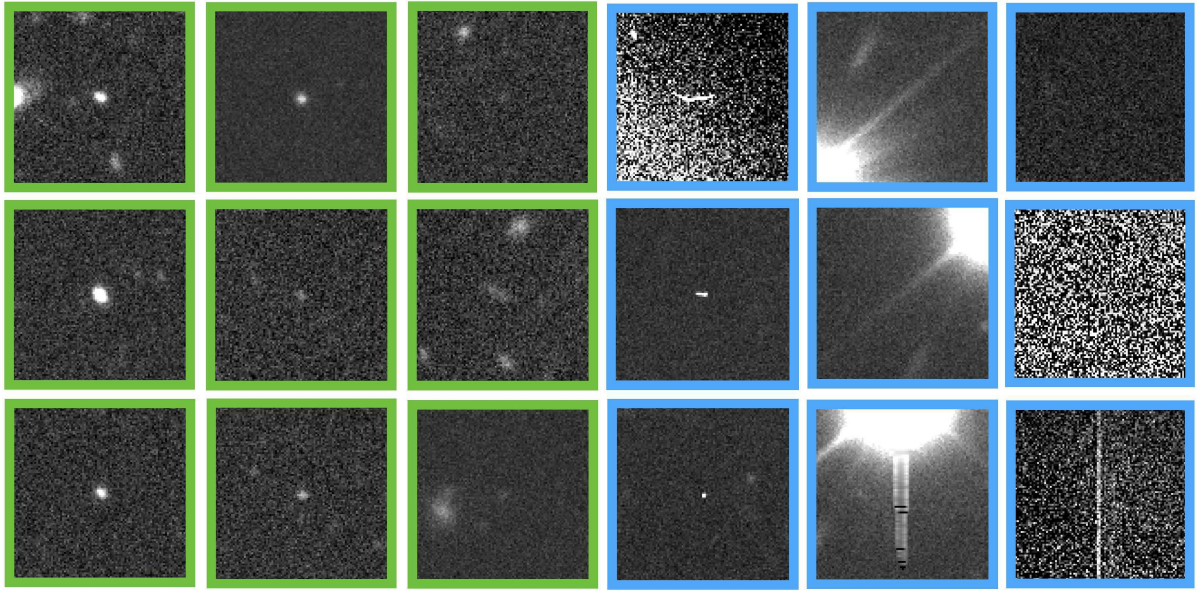} 
\caption{Samples of real (green) and bogus (blue) detections that were identified by the real-bougs system. The size of each image is roughly consist with 17'' by 17'' }\label{fig:7}
\end{figure}

The feature distributions suggest that our real-bogus system works properly. The next-step is examining of the detection images to understand that if the system can really separate the real and bogus detections. Figure~\ref{fig:7} shows the real and bogus detections selected by our system. The images with green frames represent the real detections. Many of the real detections have the elongate shapes and potentially are the asteroids. The images with blue frames are some samples of the bogus detections. They are the comsic-rays, bright star spikes, pixels with interpretation or extreme low signal sources/non-detections.

\begin{figure}
\includegraphics[width = 1.\textwidth]{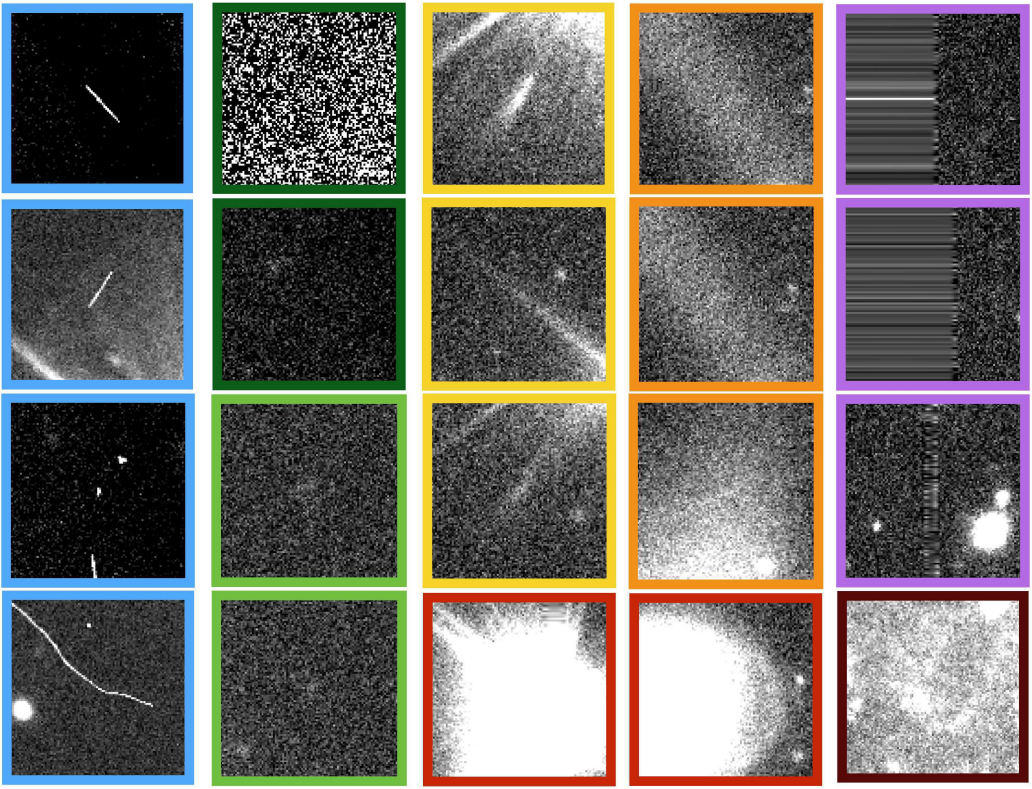} 
\caption{The samples of the real-bogus system identified bogus sources, but without ``bad flags''. They can be classified as 1. cosmic-rays (blue), 2. non-detections (deep green), 3. bright star spikes (yellow), 4. optical effect (orange), 5. over-exposure bright sources (red), 6. interpretation pixels (purple), 7. 1. faint blurry sources (light green) and extended sources in crowded-field (brown). }\label{fig:8}
\end{figure}

We therefore examine the bogus sources without ``bad flags''. The results show in Figure~\ref{fig:8}. We found that the real-bogus system have higher sensitivity with 1. cosmic-rays (blue), 2. non-detections (deep green), 3. bright star spikes (yellow), 4. optical effect (orange), 5. over-exposure bright sources (red) and 6. interpretation pixels (purple), but could mis-classify 1. faint fussy sources (light green) and 2. extended sources in crowded-field (brown) as the bogus sources. The main purpose of our real-bogus system is to reduce the false-positive rate for the HSC-SSP Moving Object Detecting Pipeline \citep{che17}. Therefore, it will be fine for excluding some extended real sources.

\subsection{Real Case Evaluation} \label{rce}

Finally, it is important to know that how many sources will be input into the moving object detection pipeline with or without the real-bogus system. Here we use a {\tt HEALPix} for evaluation. There are total 619,207 non-stationary detections in the {\tt HEALPix} for one SSP observation run. Without any false-positive cut, the moving object detection pipeline has to process more than 0.6 million detections in the {\tt HEALPix}. With the real-bogus system, 244,407 potentially false detections are rejected, and 60$\%$ of the detections need to be processed. By comparison, using the ``bad flags'' can only reject 126,739 potentially false detections in this case.

We examined some of the remaining 0.24 million non-stationary detections and found that they are mainly the very faint real stars and galaxies. Notice that our stationary source has to have corresponding detections in different exposures within 0.5 arcsec radius. Since the detectability of very faint objects is highly sensitive to image quality (e.g., seeing and transparency), some of them cannot be removed by the stationary selection criteria. The over-exposure bright stars also contrubtion part of the real non-stationary detections. That may cause by the large centering error of the over-exposure sources. Except the stars and galaxies, the remaining non-stationary detections are dominated by the asteroids and the non-removed cosmic-rays.

\section{Discussion}

The ROC curves show that our RF real-bogus system works very effectively, thanks to many features provided by {\tt hscPipe} and the idea of that using stationary sources as the real training set. Although using the stationary sources can easily produce an uniform, large-sized training data, there is a known issue with this idea. The fast-moving asteroids do not have PSF-like shapes; they will make the ``streaks'' in the images, and the profile should be a moving PSF. \citet{ver12} and \citet{lin15} have more detailed description of the fast-moving asteroid profiles. As the result, the fast moving main-belt or near Earth asteroids could be excluded by our real-bogus system. However, we also noticed that the moving object pipeline can find a lot of asteroids with short-streaked shapes, and this result implies that the fast-moving asteroids will slip through the real-bogus system. Nevertheless, currently we have not tested the performance of the real-bogus system with streak-shaped detections. In the near future, once we tune the moving object pipeline to search for fast movers, test the real-bogus system for streak-shape detections will be listed as the highest priority in our to-do list.

In general, the IF classifier is not as effective as RF classifier for real-bogus separation, but it still has an advantage. Using the IF classifier does not require any bogus training set, it could be very useful for some surveys, for which the false detections can not be selected easily. Once the stationary sources can be identified, the IF classifier will work for the bogus rejection.

Our concept of this paper, which is using the stationary sources to produce the decent training sample and train an ML classifier to identify the real and bogus detections, can be easily applied to any sky surveys: The widely used {\tt SEXtractor} \citep{ber96} can also provide all kinds of photometry and shape measurements, which is an ideal alternative to the {\tt hscPipe} for feature extraction. Therefore the real-bogus systems can be applied widely in every asteroid survey and greatly reduce the computation time for asteroid discoveries.

Although our system and concept are designed to work for the normal, non-differential source catalogs, it is still worth to try it on the differential source catalogs, which are generated from the difference images. Once the differential source catalogs have same photometry measurements and shape moments with the normal source catalogs, this system could also work and reject the false optical transients in the differential source catalogs. Thus, for the case similar to \citet{mor16}, we may be able to avoid the extra work for building the synthetic transients for training.

\section{Summary}

We built a real-bogus system to reduce the false-positive rates for the non-differential source catalog based HSC-SSP Moving Object Detecting Pipeline. The ML-classifier based system was trained by the stationary sources as the real data-set. The bogus set was selected by the ``bad flags'' generated by {\tt hscPipe}. We used 47 features, which are all provided by {\tt hscPipe}, for the classifier to separate the real and bogus detections. Most of the features are correlated to photometry measurements and shape moments.

We tested two different machine learning classifiers, 1. the supervised random forest (RF), and 2. the unsupervised isolation forest (IF), both of them can have a result of the decent separation of real and bogus detections. However, the RF classifier outperforms IF classifier; RF can reach $99\%$ of tpr at $5\%$ of fpr, by contrast, the IF can only reach a tpr $\sim 90\%$ at a fpr $= 15\%$. Therefore, we use RF for the real-bogus separation. 

Our real-bogus system is twice as effective as applying ``bad flags'' to reject the false positives; in our test the real-bogus system rejected more than 0.24 of 0.6 million non-stationary detections in a {\tt HEALPix}, but the ``bad flags'' can only reject less than 0.12 million detections.

Finally, we suggest that the concept of real-bogus system that training by stationary sources can be applied to any modern source catalogs based asteroid surveys to reduce the false-positive rates and improve the system performance.

\begin{ack}
We are grateful to Paul Price and Hisanori Furusawa for kindly suggestions and helps in data process.

This work was supported in part by MOST Grant: MOST 104-2119-008-024 (TANGO II) and MOE under the Aim for Top University Program NCU, and Macau Technical Fund: 017/2014/A1 and 039/2013/A2. HWL acknowledges the support of the CAS Fellowship for Taiwan-Youth-Visiting-Scholars under the grant no. 2015TW2JB0001.

The Hyper Suprime-Cam (HSC) collaboration includes the astronomical communities of Japan and Taiwan, and Princeton University. The HSC instrumentation and software were developed by the National Astronomical Observatory of Japan (NAOJ), the Kavli Institute for the Physics and Mathematics of the Universe (Kavli IPMU), the University of Tokyo, the High Energy Accelerator Research Organization (KEK), the Academia Sinica Institute for Astronomy and Astrophysics in Taiwan (ASIAA), and Princeton University. Funding was contributed by the FIRST program from Japanese Cabinet Office, the Ministry of Education, Culture, Sports, Science and Technology (MEXT), the Japan Society for the Promotion of Science (JSPS), Japan Science and Technology Agency (JST), the Toray Science Foundation, NAOJ, Kavli IPMU, KEK, ASIAA, and Princeton University. 

This paper makes use of software developed for the Large Synoptic Survey Telescope. We thank the LSST Project for making their code available as free software at  http://dm.lsst.org

The Pan-STARRS1 Surveys (PS1) have been made possible through contributions of the Institute for Astronomy, the University of Hawaii, the Pan-STARRS Project Office, the Max-Planck Society and its participating institutes, the Max Planck Institute for Astronomy, Heidelberg and the Max Planck Institute for Extraterrestrial Physics, Garching, The Johns Hopkins University, Durham University, the University of Edinburgh, Queen’s University Belfast, the Harvard-Smithsonian Center for Astrophysics, the Las Cumbres Observatory Global Telescope Network Incorporated, the National Central University of Taiwan, the Space Telescope Science Institute, the National Aeronautics and Space Administration under Grant No. NNX08AR22G issued through the Planetary Science Division of the NASA Science Mission Directorate, the National Science Foundation under Grant No. AST-1238877, the University of Maryland, and Eotvos Lorand University (ELTE) and the Los Alamos National Laboratory.

Based on data collected at the Subaru Telescope and retrieved from the HSC data archive system, which is operated by Subaru Telescope and Astronomy Data Center, National Astronomical Observatory of Japan.

\end{ack}


\begin{thebibliography}{}
% Journals(e.g. A\&A,ApJ,AJ,NMRAS,PASP ...)
% Authors, Year, Journal, Vol#, Page#
% Journal Title Abbreviation >> http://www.asj.or.jp/pasj/Jabb.html




\bibitem[Armstrong et al.(2017, in~prep)]{arm17} Armstrong, R., Bickerton, S., Bosch, J., et al. 2017 \pasj, in prepare

\bibitem[Bailey et al.(2007)]{bai07} Bailey, S., Aragon, C., Romano, R., et al.\ 2007, \apj, 665, 1246 


\bibitem[Baron \& Poznanski(2017)]{bar17} Baron, D., \& Poznanski, D.\ 2017, \mnras, 465, 4530 

\bibitem[Bernstein \& Jarvis(2002)]{ber02} Bernstein, G.~M., \& Jarvis, M.\ 2002, \aj, 123, 583 

\bibitem[Bertin \& Arnouts(1996)]{ber96} Bertin, E., \& Arnouts, S.\ 1996, \aaps, 117, 393 


\bibitem[Bloom et al.(2012)]{blo12} Bloom, J.~S., Richards, J.~W., Nugent, P.~E., et al.\ 2012, \pasp, 124, 1175 

\bibitem[Breiman, L. (2001)]{bre01} Breiman, L. Random forests. Machine learning, 45, 5-32.

\bibitem[Bosch et al.(2017, in~prep)]{bos17} Bosch et al. 2017, \pasj, in prepare

\bibitem[Brink et al.(2013)]{bri13} Brink, H., Richards, J.~W., Poznanski, D., et al.\ 2013, \mnras, 435, 1047 

\bibitem[Cavuoti et al.(2017)]{cav17} Cavuoti, S., Amaro, V., Brescia, M., et al.\ 2017, \mnras, 465, 1959 

\bibitem[Cavuoti et al.(2015)]{cav15} Cavuoti, S., Brescia, M., Tortora, C., et al.\ 2015, \mnras, 452, 3100 

\bibitem[Chen et al.(2017,~in~prep)]{che17} Chen et al. 2017, \pasj, in prepare

\bibitem[du Buisson et al.(2015)]{bui15} du Buisson, L., Sivanandam, N., Bassett, B.~A., \& Smith, M.\ 2015, \mnras, 454, 2026 

\bibitem[Gerdes et al.(2010)]{ger10} Gerdes, D.~W., Sypniewski, A.~J., McKay, T.~A., et al.\ 2010, \apj, 715, 823 

\bibitem[Goldstein et al.(2015)]{gol15} Goldstein, D.~A., D'Andrea, C.~B., Fischer, J.~A., et al.\ 2015, \aj, 150, 82 

\bibitem[G{\'o}rski et al.(2005)]{gor05} G{\'o}rski, K.~M., Hivon, E., Banday, A.~J., et al.\ 2005, \apj, 622, 759 

\bibitem[Hirata \& Seljak(2003)]{hir03} Hirata, C., \& Seljak, U.\ 2003, \mnras, 343, 459 

\bibitem[Huppenkothen et al.(2017)]{hup17} Huppenkothen, D., Heil, L.~M., Hogg, D.~W., \& Mueller, A.\ 2017, \mnras, 466, 2364 

\bibitem[Kawanomoto et al.(2017,~in~prep))]{kaw17} Kawanomoto et al. 2017, \pasj, in prepare

\bibitem[Komiyama et al.(2017,~in~prep)]{kom17} Komiyama et al. 2017, \pasj, in prepare

\bibitem[Krone-Martins et al.(2014)]{kro14} Krone-Martins, A., Ishida, E.~E.~O., \& de Souza, R.~S.\ 2014, \mnras, 443, L34 


\bibitem[Lin et al.(2015)]{lin15} Lin, H.~W., Yoshida, F., Chen, Y.~T., Ip, W.~H., \& Chang, C.~K.\ 2015, Icarus, 254, 202 

\bibitem[Liu et al.(2008)]{liu08} Liu, F. ~T., Ting, K.~M., \& Zhou, Z.~H. In Data Mining, 2008. ICDM'08. Eighth IEEE International Conference on pp. 413-422. IEEE.


\bibitem[Lochner et al.(2016)]{loc16} Lochner, M., McEwen, J.~D., Peiris, H.~V., Lahav, O., \& Winter, M.~K.\ 2016, \apjs, 225, 31 


\bibitem[Masci et al.(2017)]{mas17} Masci, F.~J., Laher, R.~R., Rebbapragada, U.~D., et al.\ 2017, \pasp, 129, 014002 

\bibitem[Miller et al.(2017)]{mil17} Miller, A.~A., Kulkarni, M.~K., Cao, Y., et al.\ 2017, \aj, 153, 73 

\bibitem[Miller et al.(2015)]{mil15} Miller, A.~A., Bloom, J.~S., Richards, J.~W., et al.\ 2015, \apj, 798, 122 

\bibitem[Miyazaki et al.(2017,~in~prep)]{miy17} Miyazaki et al. 2017, \pasj, in prepare

\bibitem[Morii et al.(2016)]{mor16} Morii, M., Ikeda, S., Tominaga, N., et al.\ 2016, \pasj, 68, 104 


\bibitem[Ostrovski et al.(2017)]{ost17} Ostrovski, F., McMahon, R.~G., Connolly, A.~J., et al.\ 2017, \mnras, 465, 4325 

\bibitem[Pedregosa et al. (2011)]{ped11} Pedregosa, F., Varoquaux, G., Gramfort, A., et al.\ 2011, Journal of Machine Learning Research, 2825-2830.
ISO 690	


\bibitem[Rowe et al.(2015)]{row15} Rowe, B.~T.~P., Jarvis, M., Mandelbaum, R., et al.\ 2015, Astronomy and Computing, 10, 121 


\bibitem[Sadeh et al.(2016)]{sad16} Sadeh, I., Abdalla, F.~B., \& Lahav, O.\ 2016, \pasp, 128, 104502 


\bibitem[Samui \& Samui Pal(2017)]{sam17} Samui, S., \& Samui Pal, S.\ 2017, NewAstronomy, 51, 169 

\bibitem[Takata et al.(2017,~in~prep)]{tak17} Takata et al. 2017, \pasj, in prepare

\bibitem[Vere{\v s} et al.(2012)]{ver12} Vere{\v s}, P., Jedicke, R., Denneau, L., et al.\ 2012, \pasp, 124, 1197 



\bibitem[Waszczak et al.(2017)]{was17} Waszczak, A., Prince, T.~A., Laher, R., et al.\ 2017, \pasp, 129, 034402 

\bibitem[Wolf et al.(2017)]{wol17} Wolf, C., Johnson, A.~S., Bilicki, M., et al.\ 2017, \mnras, 466, 1582 


\bibitem[Wright et al.(2015)]{wri15} Wright, D.~E., Smartt, S.~J., Smith, K.~W., et al.\ 2015, \mnras, 449, 451 

\bibitem[Zheng \& Zhang(2012)]{zhe12} Zheng, H., \& Zhang, Y.\ 2012, \procspie, 8451, 845133 




\end{thebibliography}
\end{document}